\documentstyle[preprint,prl,aps]{revtex}
\begin{document}
\draft
\title{ The Hofstadter Spectrum and Photoluminescence}
\author{Lian Zheng and H. A. Fertig}
\address{Department of Physics and Astronomy,
University of Kentucky, Lexington, Kentucky 40506}
\date{\today}
\maketitle
\begin{abstract}
The observability of the Hofstadter spectrum
generated by a Wigner crystal
using photoluminescence
techniques is studied.
Itinerant hole geometries are examined, in which a
hole may combine directly
with electrons in the lattice.
It is found that when the effect of lattice
distortions of the WC due to interactions with
the hole are accounted for, only the largest
Hofstadter gaps are observable.
To overcome the problems of lattice distortion,
a novel geometry is proposed, involving
a two layer system with electrons in one layer
forming a WC and in the other a full Landau
level.   It is found that recombination of electrons in the
full Landau level with {\it localized} holes reflects
the full Hofstadter spectrum of the lattice.
\end{abstract}
\narrowtext
An electron gas is expected to condense into a Wigner crystal \cite{wc}(WC)
below some critical density.
In the absence of a magnetic field,
this condensation occurs
when the cost in kinetic energy
due to crystallization
is outweighed by the decrease in Coulomb energy.
For a two-dimensional electron system (2DES),
the kinetic energy
scales like $K=\hbar^2/m^*a_o^2$, while the Coulomb energy
scales like $V=e^2/\epsilon a_o$, where $a_o$ is the mean inter-electron
distance and $\epsilon$ is the dielectric constant of the host
material, and $m^*$ and $e$ are the electron mass and charge respectively.
The relevant parameter is the ratio $r_s=V/K=a_o/a_B$, where
$a_B=\hbar^2\epsilon/m^*e^2$ is the Bohr radius.
Monte Carlo simulation \cite{mc}
predicts that a 2DES crystallizes
for $r_s\geq37$.
When a strong magnetic field is applied perpendicular
to the 2DES, the situation is changed qualitatively,
as the kinetic energy is quenched into discrete Landau levels, and
the zero-point fluctuations in the lowest Landau level
are confined within a magnetic length $l_o=(\hbar c/eB)^{1/2}$,
where $c$ is the speed of light and $B$ is the applied
magnetic field.
Once $l_o$ is sufficiently small compared to the typical
inter-particle distance
$a_o$, crystallization occurs. The ratio $l_o/a_o$
can be characterized by the Landau level filling factor
$\nu=2\pi l_o^2n$, where $n$ is the density of the 2DES.
Crystallization will occur for
sufficiently small $\nu$
for any given density.
Early theoretical estimates \cite{gm}
put the critical filling factor of crystallization
at about $\nu_c\sim 1/6.5$.
Recent numerical calculations \cite{pla} suggest that
larger values of the critical filling factor are possible
by considering the Landau level mixing.

Crystallization in the absence of a magnetic field has been
observed \cite{hel,mos} for
electrons trapped on the surface
of liquid helium.
The magnetic-field-induced WC
has not been unambiguously observed.
Reentrant insulating phases around $\nu=1/5$
for a 2DES \cite{fif}
and around $\nu=1/3$ for a hole system \cite{thi} have been found
and studied by various techniques.
While the assumption of a pinned WC as the ground state of the insulating
phase is found to be qualitatively consistent with many of the experimental
results \cite{fif,thi,tra,rf,rf2,cr,pl,pl2},
alternative interpretations \cite{slz,kli}
based on disorder-induced states are not ruled out.
Despite increased efforts in recent years,
the nature of the insulating phase still remains
unclear. One of the experiments which may help to
resolve this puzzle is the photoluminescence (PL)
from the recombination of the electrons in the insulating phase
with nearby holes.
It has been shown\cite{lfd} that PL spectra can yield
information about the single particle density of states
for both the holes and the recombining electrons.
This means that
PL potentially contains important information
about the electronic state, because a charged
particle subject both to a periodic potential
due to an electron lattice and a
magnetic field has a characteristic energy spectrum.
This so-called Hofstadter spectrum\cite{hof}
consists of a series of bands and gaps
that is very sensitive to the number of
magnetic flux quanta $\phi/\phi_0$ passing
through each unit cell of the lattice: for $\phi/\phi_0=p/q$,
with $p$ and $q$ integers, there are $p$ bands and $p-1$
gaps.  It has been proposed\cite{lfd} that
itinerant hole PL experiments, in which some of the
holes are thermally excited to higher Hofstadter bands, offers
a possible avenue for observing the lowest few
bands of this spectrum.
Because of their great sensitivity to the
magnetic field, successfully identifying these bands from
the PL spectrum would provide convincing
evidence for the existence of a WC.

In this work, we shall study the relevant densities of
states (DOS) that enter into the PL spectrum.
We begin with the case of itinerant
holes interacting with a nearby electron crystal.
We will argue that
most of the gaps in the Hofstadter spectrum
are greatly suppressed, due to the
lattice distortion caused by the electron-hole interaction.
Only the energy levels corresponding to
the localized orbitals of a hole trapped at a single lattice electron
are likely to be observed in PL experiments.
The energies of these levels are
computed, and
their relationship with
the Hofstadter spectrum of a
system without lattice distortions is discussed.
We shall then propose a novel geometry
involving a double-layer structure which avoids all the
difficulties introduced by the lattice distortions,
and provides a great potential to observe
the unperturbed Hofstadter spectrum associated with
the electron Wigner crystal.

Our results for itinerant holes may be
summarized as follows.  Firstly,
the hole plane must be at least
a minimum distance $d > d_{tr} \approx 0.45a_0$ away from the WC
to avoid inducing the formation of an interstitial
defect in the lattice\cite{pin}.
For $d$ larger than this critical value,
the electron-hole potential still induces
lattice distortions, although they are
considerably weaker than for small $d$.
It is found
that only the largest gaps in the Hofstadter spectrum
actually survive these weak distortions, and that
the splittings among the PL lines from
holes thermally excited to higher energy states
are very close to, and are
best understood in terms of, the spectrum
of a hole bound to a single electron in the lattice.
In this situation, one expects to see a single new
line in the hole DOS as the flux per unit cell (or, equivalently,
per electron) is increased by a single flux quantum.
Thus, lattice distortions eliminate much of the
sensitivity to the precise number of flux quanta
per unit cell seen in the full Hofstadter spectrum.
Nevertheless,
if this spectrum were observed
in actual PL experiments, this would provide strong
evidence that the electrons have localized at individual
sites, and would distinguish between a liquid
or ``Hall insulator'' ground state\cite{slz} and
one in which the individual electrons have localized\cite{klitzing}.

To analyze the expected hole DOS, we employ the
Hartree-Fock approximation.  One must find a
sensible method for dealing with the lattice
deformations (i.e., polaron effects\cite{mahan})
that necessarily accompany a hole moving near
the WC.
Within the Hartree-Fock approximation \cite{cote},
two possible states exist for the system.
One is the state where the hole has an equal possibility of
residing on each lattice site and the crystal symmetry is preserved.
The other is the state where the hole is localized at a single lattice
site and the crystal symmetry is destroyed.
We find that the
state where the hole is trapped at a single lattice site
is energetically favored over the other state.
Therefore, we shall take this case as the relevant configuration
for the system at $d>d_{tr}$,
and  study the single-particle excitation spectrum of the hole
in this configuration.

To understand the connection of this localized hole spectrum
with the Hofstadter spectrum, it is most convenient
to use a semiclassical approach\cite{wilkinson}.
We begin by considering filling factor $\nu=1/m$,
ignore lattice distortions,
and focus on the lowest few states.  Semiclassically,
there is a set of bound states of the hole
for each lattice site, characterized by constant
energy contours where the wavefunctions are
maximized,
which represent classical drift orbits for the
hole in the potential of the electron lattice.
The allowed quantum states associated with these orbits are
determined by the requirement that
each state's orbit must contain approximately one
more flux quantum than the state immediately below it\cite{fertig}.
Bound states in different unit cells
enclosing a given number of flux quanta are
coupled together via tunneling to form the Hofstadter
bands.
The tunneling matrix element
between sites for the low-lying states scales\cite{fertig}
as $e^{-\alpha a_0^2/l_0^2}$,
where $\alpha$ is a number of order unity that depends on
the precise shape of the potential.  It should be kept in mind that in the
low filling factor limit, this is quite small, so that
the bands will be very narrow.

When the filling factor is changed slightly away from
$\nu=1/m$ to $\nu=p/q$, the magnetic field introduces an
extra phase that must be accumulated when the hole tunnels
between nearest neighboring sites. This breaks up each
band near the bottom of the spectrum into
$p$ separate bands\cite{wilkinson}.  However,
because these gaps arise due to
phases introduced in tunneling, the gaps
themselves also scale as $e^{-\alpha a_0^2/l_0^2}$, which
is quite small. This is in accordance with more exact
calculations of the Hofstadter spectrum, for small
filling factors\cite{hof}.

The result is that the Hofstadter
gaps based on the bound state spectrum of the holes
will be relatively
easy to observe\cite{rk1} (these gaps are typically of
order of several degrees $K$), whereas the smaller gaps based
on tunneling between sites are quite difficult
to resolve.  The effect of lattice distortions
greatly compounds this.
Even with lattice distortions, {\it in principle} the hole
can tunnel between lattice sites, although in doing so
the lattice distortion around the initial site must relax
and a new lattice distortion must develop around the
final site.  This significantly reduces the effective tunneling
matrix element between sites for the hole.  With such
small effective tunneling amplitudes, for
experimentally relevant quantities it may be
neglected.  Thus, only the largest gaps in the
Hofstadter spectrum should be observable in these
geometries, and for these low-lying levels they
are to an excellent approximation given by
the bound state spectrum of an individual lattice site.
The Hartree-Fock state in which the lattice
deforms around a localized hole should give an
excellent approximation for the effects of lattice
distortion on these lowest few states.

In order to apply the Hartree-Fock approximation \cite{cote}
to the present case where the crystal symmetries no longer exist,
we use a super-cell technique,
{\it i.e.}, we consider a periodic array of
holes, each  trapped at a single lattice site.
The super-lattice has primary lattice vectors which are
$k$ times as large as the lattice vectors of the original
electron crystal,
where $k$ is an integer.
In this way we are dealing with a hole-electron double-layer system.
These two layers are coupled through an electrostatic potential
$H_{h-e}=-(1/\Omega)\sum_{\bf G}
(2\pi e^2/\epsilon G)e^{-Gd}\rho_h({\bf G})
\rho_e(-{\bf G})$, where $\Omega$ is the area of the system
and $\rho_h({\bf G})$ and $\rho_e({\bf G})$ are respectively
the hole and electron densities.
For simplicity the hole-hole interaction may be ignored,
because in practice the hole density is
much smaller than the electron density.
The equations from this Hartree-Fock approximation
are then solved numerically
at zero-temperature
and under the condition that
only the lowest Landau level is occupied.

The density of states of the holes
is shown in Fig. (\ref{fig1}). It was calculated with
an electron Landau level filling factor $\nu=1/6$,
a hole setback distance of $d=0.7a_o$, and
with 9 electrons per super-cell.
One can see that, due to the deformation of the electron crystal,
the density of states differs greatly from
the 6-band spectrum expected in the case of an unperturbed crystal.
The result of Fig. (\ref{fig1}) can be understood as
the bound state spectrum for the hole in the potential
well caused by the lattice deformation.
We note that, while this density of states
looks quite different than the Hofstadter spectrum,
the energy differences between the low-lying energy levels are almost
identical to the separations
between the lowest few Hofstadter bands of the
undistorted lattice.  This just reflects the fact
that the large gaps in the Hofstadter spectrum are
essentially those of the spectrum for
a hole bound to a single lattice site.
The several lowest levels in the excitation spectrum of Fig. (\ref{fig1}),
which may be observable in experiment, are
marked by the dotted lines.
Figs. ~\ref{fig2} and ~\ref{fig3} illustrate the
evolution of these energy levels as a function
of $d$ and $B$.

As can be seen from the above discussion,
there are several limitations
associated with the itinerant hole experiments.
Lattice distortions collapse all but
the largest
Hofstadter
gaps, leaving a spectrum that is best understood
as the bound states of a hole with a single lattice site.
Even among these states,
only the lowest few are expected
to be observed in the actual PL spectrum.
There are two reasons for this:  higher states
of the holes are only occupied due to thermal fluctuations,
and those holes
which do occupy such states tend to be physically distant
from the electrons, so that the resulting PL power will
be small\cite{c1}.  Clearly, it is highly desirable to
find a situation in which the Hofstadter spectrum
may be observed unperturbed, in its entirety.

A geometry that overcomes essentially all these problems
and provides a great potential to observe
the unperturbed Hofstadter spectrum associated with
the electron Wigner crystal
is illustrated in Fig. ~\ref{fig0}.   It is a
system of two layers, one with low enough density that the electrons
would be expected to crystallize (we will call
this the ``crystal layer''), the other at much
higher density, enough to fill an integral number of
Landau levels (for concreteness, we will take $\nu=1$,
and call this the ``probe layer.'')
A layer of acceptors
may then be grown close to the $\nu=1$ layer.  The
idea is to observe PL from recombination processes
between the $\nu=1$ layer and {\it localized} holes in the cores of
the acceptors.  In this geometry, PL is a measure of
the density of states for the $\nu=1$ electron layer.
Because there is a large gap to excited states of
this layer, it is easily shown that its density
must be uniform, so that the probe layer cannot
affect the electronic structure of the crystal layer.
Nevertheless, one may compute the density of single
particle states for the probe layer, and it is easily
shown
that it is {\it identical} to that of a single
(negatively) charged particle moving in the periodic
potential due to the crystal layer.

That the probe layer density of states
reflects the Hofstadter spectrum may be
seen most easily in the Hartree-Fock
approximation\cite{cote}.  The retarded Green's function for
electrons in the probe layer satisfy the
equation
$$
(\omega + i\delta +\mu)G_p({\bf G},\omega)
-U(0)G_p({\bf G},\omega)
-\sum_{{\bf G^{\prime}}} V({\bf G}- {\bf G^{\prime}})
e^{i {\bf G} \times {\bf G^{\prime}} l_o^2/2}
G_p({\bf G^{\prime}},\omega)
=\delta_{{\bf G},0},
$$
where ${\bf G}$ are the reciprocal lattice vectors of
the WC, $G_p({\bf G},\omega)$ is the Green's
function for electrons in the probe layer,
$U({\bf G}) \equiv W({\bf G})<\rho_p({\bf G})>$,
$W$ is the difference of the direct and exchange interactions,
and $V$ is the periodic potential due to the WC in which
the electrons in the probe layer move.  The key insight
is that because the probe layer is a precisely filled
Landau level, if we ignore mixing of higher Landau levels,
the electron density in the probe layer must be uniform.
This means that
$U({\bf G})$ is only different from zero for ${\bf G}=0$,
so that the only effect of electron-electron interactions
within the probe layer is to shift the chemical potential
$\mu$ by an overall constant.  Thus, $G_p$ obeys {\it precisely} the
equation of motion for a single electron in a periodic potential.
Finally, observing that the density of states $D(E)$ is just
proportional to Im$G({\bf G}=0,E)$, one sees that
this is identical to the result expected for a single
electron in a periodic potential.

There are several advantages to this geometry over
presently available ones:  (1) The lattice is not
distorted either by the core holes (which are screened
in their initial state\cite{pl})
or by the probe layer, so that the Hofstadter spectrum
is limited only by the intrinsic disorder of the sample,
not by the probes themselves.  None of
the Hofstadter gaps are collapsed by lattice distortions.
(2) Unlike the itinerant hole experiments, {\it all} the
Hofstadter bands are filled in the probe layer (which
is why it has a net uniform density), so that all the
bands can contribute to the PL.  (3) Since the density
of the probe layer is uniform, and the holes are
placed randomly in the plane, the electrons in
{\it all} the Hofstadter states
will have a significant overlap with the holes, not
just the lowest few.  (4) Since the geometry
employs localized holes, there are no ambiguities
as to the initial state of the holes, which has
made presently available valence band hole data\cite{pl2}
difficult to interpret. (5) Unlike present localized hole
(and valence band hole)
experiments, one is not removing an electron from
the crystal layer, so shakeup effects\cite{shakeup}
should not be very pronounced; this will especially
be true if the acceptors are close to the probe
layer.  We believe this
geometry takes advantage of the best features of
both itinerant carrier and localized hole experiments,
and has an excellent potential to probe
the Hofstadter spectrum.

In summary, we have studied the single-particle density of states
for a hole near an electron WC with the Hartree-Fock
approximation.
The holes are not bounded to any impurity or disorder sites.
We take into consideration the perturbation of the crystal
by the holes.
We have found that the
low-lying excitations for the holes
at $d>d_{tr}=0.45a_o$
consist of well-separated energy levels.
We have evaluated the energy splittings for these levels
at different distance $d$ and for
different magnetic fields.
We suggest that these low-lying levels can produce
additional lines in the photoluminescence spectrum
of the hole-WC system
under experimentally relevant conditions.
A novel geometry using a full Landau level as a probe
layer is proposed to overcome the problems
intrinsic to itinerant hole geometries.
This geometry provides a great potential to observe
the unperturbed Hofstadter spectrum associated with
the electron Wigner crystal.

This work is supported by the National Science Foundation through Grant
No. DMR-9202255.  HAF acknowledges support from the
Sloan Foundation and the Research Corporation.

\begin{figure}
\caption{
Density of states for the holes at a setback distance
of $d/a_o=0.7$ and at a filling factor of $\nu=1/6$.
The lowest five levels, marked by $\epsilon_i$ for $i=1\sim5$,
are well-separated in energy. Each of the five levels
has a degeneracy equal to the number of the holes in the system.
}
\label{fig1}
\end{figure}

\begin{figure}
\caption{
The energies of the four lowest excitation levels
of the holes $\epsilon_i$ ($i=2\sim5$)
as functions of the hole setback distance $d$
at a filling factor of $\nu=1/6$.
The energies are with respect to the energy of the
lowest level $\epsilon_1$.
}
\label{fig2}
\end{figure}

\begin{figure}
\caption{
The energies of the four lowest excitation levels
of the holes $\epsilon_i$ ($i=2\sim5$)
as functions of magnetic field strength at a hole
setback distance of $d/a_o=0.7$.
The energies are with respect to the energy of the
lowest level $\epsilon_1$.
}
\label{fig3}
\end{figure}

\begin{figure}
\caption{
Proposed two layer geometry for observing the
Hofstadter spectrum of a Wigner crystal.
}
\label{fig0}
\end{figure}

\begin{references}
\bibitem{rk1}
Energy level broadening by disorder scattering and so on, which
are not taken into account in this work, must be smaller than
the energy splittings in Figs.~\ref{fig2}
and ~\ref{fig3} in order to observed separated lines
from the PL data.
\bibitem{c1} However, estimates of the PL power from the lowest
few Hofstadter bands indicate that they should be observable.
See Ref. \cite{lfd}.
\end{references}
\end{document}